\documentclass[%
 reprint,%
 amsmath,%
 amssymb,%
 aps,%
 floatfix,
 superscriptaddress,
 prl]{revtex4-2}

\usepackage{graphicx}
\graphicspath{{./figs}}
\usepackage{mathtools}
\usepackage{amsmath, amssymb}
\usepackage{dsfont}
\usepackage{bm}
\usepackage[dvipsnames]{xcolor}
\usepackage[colorlinks,%
            pdfusetitle,%
            bookmarksopen=false,%
            urlcolor=Blue,%
            citecolor=Blue,%
            linkcolor=Blue,%
            pdfencoding=auto,%
            psdextra]{hyperref}
\usepackage{physics}
\usepackage{upgreek}
\usepackage{bbm}
\usepackage[normalem]{ulem}
\usepackage{siunitx}
\usepackage{booktabs}
\usepackage{leftindex}

\usepackage{soul}

\newcommand{\trippzero}{\ensuremath{\leftindex^3{P}_0}}

\newcommand{\singszero}{\ensuremath{\leftindex^1{S}_0}}

\newcommand{\tripsone}{\ensuremath{\leftindex^3{S}_1}}

\newcommand{\trippzot}{\ensuremath{\leftindex^3{P}_{0,1,2}}}

\newcommand{\aitime}{\ensuremath{31(2)\,\si{\nano\second}}}
\newcommand{\barecontrast}{\ensuremath{0.970(4)}}
\newcommand{\aicontrast}{\ensuremath{0.977(4)}}
\newcommand{\aishift}
{\ensuremath{235.6(1)\,\si{\hertz}}}
\newcommand{\aishiftfom}
{\ensuremath{2\pi \times 7.3(5)\, \si{\micro rad}}}

\begin{document}

\setstcolor{red}

\title{Autoionization-enhanced Rydberg dressing by fast contaminant removal}

\author{Alec Cao}
\author{Theodor \surname{Lukin Yelin}}
\author{William J. Eckner}
\author{Nelson \surname{Darkwah Oppong}}
\author{Adam M. Kaufman}
\affiliation{%
JILA, University of Colorado and National Institute of Standards and Technology,
and Department of Physics, University of Colorado, Boulder, Colorado 80309, USA
}%


\begin{abstract}
    Rydberg dressing is a powerful tool for entanglement generation in long-lived atomic states. 
    While already employed effectively in several demonstrations, a key challenge for this technique is the collective loss triggered by blackbody-radiation-driven transitions to contaminant Rydberg states of opposite parity. 
    We demonstrate the rapid removal of such contaminants using autoionization (AI) transitions found in alkaline-earth-like atoms.
    The AI is shown to be compatible with coherent operation of an array of optical clock qubits.
    By incorporating AI pulses into a stroboscopic Rydberg dressing (SRD) sequence, we enhance lifetimes by an order of magnitude for system sizes of up to 144 atoms, while maintaining an order of magnitude larger duty cycle than previously achieved. 
    To highlight the utility of our approach, we use the AI-enhanced SRD protocol to improve the degree of spin-squeezing achieved during early-time dressing dynamics.
    These results bring Rydberg dressing lifetimes closer to fundamental limits, opening the door to previously infeasible dressing proposals.
\end{abstract}

\maketitle

Neutral atoms are a versatile platform for quantum science.
A key tool of this platform is coupling to high-lying Rydberg states for access to strong interactions~\cite{saffman2010quantum, browaeys2020many, morgado2021quantum}.
These interactions have featured prominently in recent advances for high-fidelity gates~\cite{evered2023high, ma2023high, peper2025spectroscopy, tsai2024benchmarking, radnaev2024universal}, quantum many-body simulation~\cite{ebadi2021quantum, scholl2021quantum, semeghini2021probing, chen2023continuous, scholl2023erasure, shaw2024benchmarking}, and entanglement-enhanced metrology~\cite{schine2022long,eckner2023realizing, bornet2023scalable, hines2023spin, cao2024multi, finkelstein2024universal}.
One coupling scheme which has garnered substantial theoretical interest is Rydberg dressing~\cite{bouchoule2002spin, johnson2010interactions, henkel2010three, pupillo2010strongly, honer2010collective, grusdt2013fractional, mattioli2013cluster, gil2014spin, glaetzle2014quantum, balewski2014rydberg, bijnen2015quantum, li2015exotic, glaetzle2015designing, glaetzle2017coherent, potirniche2017floquet, kaubruegger2019variational, mitra2020robust, young2023enhancing}, in which a detuned laser causes the Rydberg van der Waals potential to be effectively experienced as a softcore interaction in a long-lived state~\cite{henkel2010three, gil2014spin}.
In this way, dressing enables on-demand interactions in a high-coherence, controllable subspace.
Experimentally, Rydberg dressing has been used to implement entangling gates~\cite{jau2016entangling, schine2022long}, many-body spin dynamics~\cite{zeiher2016many, zeiher2017coherent, borish2020transverse, steinert2023spatially, eckner2023realizing} and extended Hubbard models~\cite{guardado2021quench, weckesser2024realization}.

\begin{figure}
    \includegraphics[width = \columnwidth]{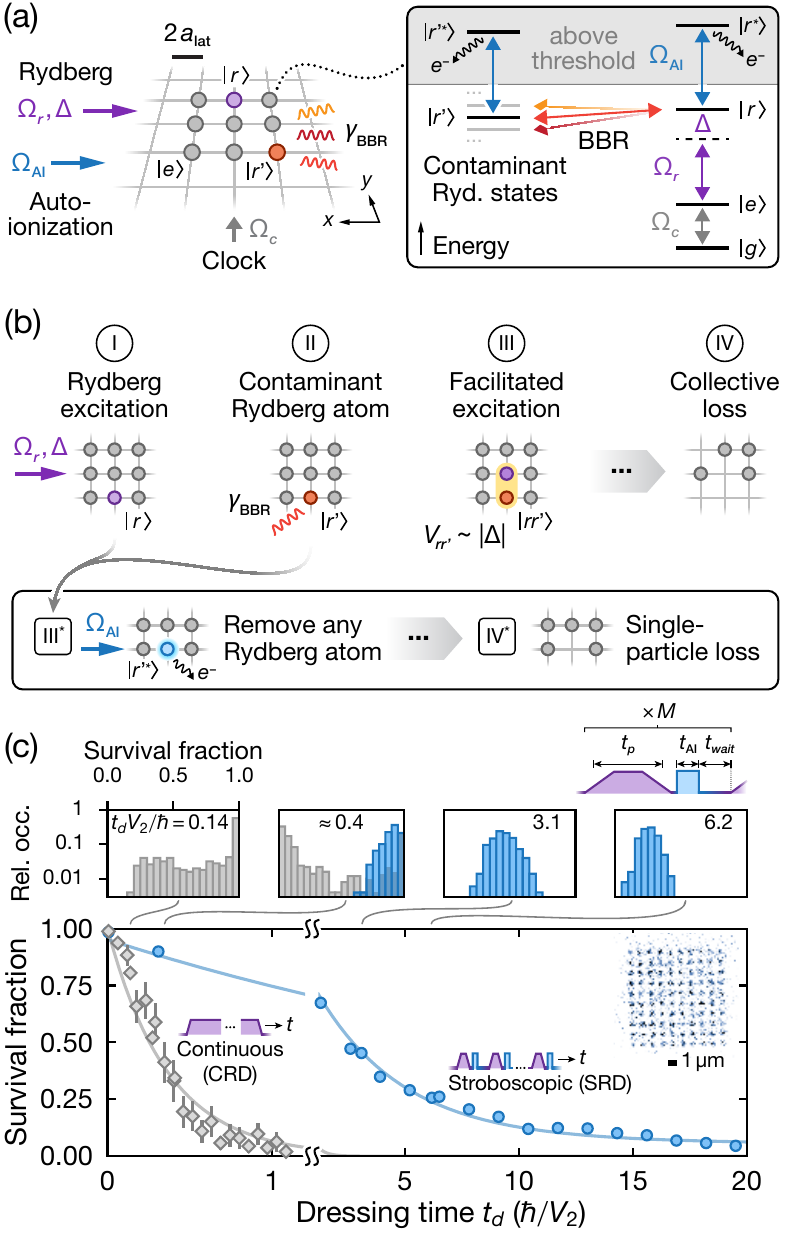}
    \caption{\label{fig:1}%
    \textbf{Suppressing collective loss in Rydberg-dressed atom arrays.}
    (a)~Schematic illustration of the experimental setup with ${{}^\mathrm{88}\mathrm{Sr}}$ atoms (circles) trapped in an optical lattice (gray lines) and lasers (colored arrows) driving the clock, Rydberg, and AI transitions.
    BBR photons ($\gamma_\mathrm{BBR}$, wiggly lines) couple the target Rydberg state~$\left| r \right\rangle$ to contaminant states~$\left| r^\prime \right\rangle$.
    (right) Electronic states, transitions, and couplings relevant for Rydberg dressing and AI.
    (b)~Collective-loss mechanism in Rydberg-dressed atom arrays separated into steps (I-IV) for visual clarity: BBR generates contaminant atoms~$\left| r^\prime \right\rangle$; this enables facilitated excitation and subsequent collective loss if the pair potential~$V_{r r^\prime}$ matches the laser detuning~$\Delta$.
    (bottom, III\textsuperscript{*}-IV\textsuperscript{*})~Removal of (contaminant) Rydberg atoms through AI suppresses the collective-loss mechanism.
    Ellipses (\ldots) indicates omitted steps such as Rydberg-atom decay.
    (c)~Survival of an~$N_\mathrm{tot}= 10\times10$ atom array after variable Rydberg dressing time $t_d$ with CRD~(gray diamonds) and SRD~(blue circles).
    Solid lines are numerical fits~\cite{supp}.
    Error bars are standard error of the mean and partly smaller than the marker size.
    Top right of main panel shows a single-shot fluorescence image.
    Upper histograms show the relative occurrence rate of observing different survival fractions at specific times, obtained from $\approx 500$ independent repetitions of the experiment.
    Top right diagram shows pulse sequence for SRD.}
\end{figure}

A major limitation for Rydberg dressing is blackbody radiation (BBR)~\cite{gallagher1979interactions}, which can trigger collective losses as shown in Fig.~\ref{fig:1}(b). 
In this process, BBR drives transitions from the target Rydberg state $\ket{r}$ to a set of nearby, opposite-parity Rydberg (contaminant) states $\ket{r'}$.
The resonant dipolar exchange interaction of $\ket{r'}$ with $\ket{r}$~\cite{mourachko1998many,anderson1998resonant} can shift the dressing laser into resonance, facilitating a proliferation of Rydberg excitations.
Signatures of collective loss have been observed in a number of Rydberg dressing experiments~\cite{zeiher2016many, festa2022blackbody, hollerith2022realizing, eckner2023realizing, hines2023spin}, as well as in Rydberg spectra of Bose condensates~\cite{goldschmidt2016anomalous,aman2016trap, boulier2017spontaneous}.
Critically, the collective loss can be triggered by only a single contaminant (Rydberg atom), causing the loss rate to grow for increasing ensemble size.

Collective loss can be mitigated by applying the Rydberg laser in a series of short pulses with sufficiently long wait times in-between~\cite{aman2016trap, boulier2017spontaneous, zeiher2017realization}.
Contaminants created during each short pulse decay back to low-lying atomic states during the wait, preventing facilitated excitation on subsequent pulses. 
The suppression of collective loss in dressing was recently demonstrated using such a stroboscopic Rydberg dressing (SRD) protocol~\cite{hines2023spin}.
Compared to continuous Rydberg dressing (CRD) with a single pulse, this SRD strategy suffers from a low duty cycle~\footnote{Independent of the collective loss issue, a low duty cycle can offer fundamental benefits when bridging the gap between the Rydberg dressing interaction energy and a slower time-scale such as tunneling in an optical lattice, as discussed and demonstrated in Ref.~\cite{weckesser2024realization}} as a result of waiting for the contaminants to decay.
A fast contaminant removal procedure can be used to overcome this limitation. 
In divalent atoms, such removal can be performed by exciting the remaining core electron to an autoionizing (AI) state~\cite{cooke1978doubly,xu1986sr}; from this state, the Rydberg electron is ejected due to its interaction with the core electron. 
Critically, both the core excitation and subsequent AI can occur orders of magnitude faster than typical contaminant state decay. 
Since the core transition frequency is relatively insensitive to the Rydberg state, a single AI laser is capable of removing all dominantly populated contaminant states, reducing experimental overhead.
Previously, AI has been used for improved Rydberg detection schemes~\cite{millen2011spectroscopy, lochead2013number, mcquillen2013imaging, madjarov2020high, ma2022universal}, and the core transition polarizability for shifting the Rydberg state energy~\cite{burgers2022controlling, pham2022coherent}.

In this Letter, we report on fast contaminant removal via AI in a Rydberg-dressed strontium atom array. 
We first show the compatibility of AI with coherent state manipulation, showing negligible impact on optical clock qubit coherence and further Rydberg excitation. 
We then integrate AI with SRD, achieving an order of magnitude suppression of collective loss for two-dimensional ensembles of up to 144 atoms; a Rydberg duty cycle near the 10\% level is maintained, compared to $< 1 \%$ realized previously without AI~\cite{hines2023spin}.
Finally, we explicitly demonstrate the utility of our approach by investigating spin-squeezing~\cite{gil2014spin,eckner2023realizing, hines2023spin}. 
In a regime with significant collective loss at the optimal squeezing time, we recover 3\,dB of Wineland spin-squeezing with AI-enhanced SRD while suppressing around 5\,dB of noise relative to CRD.
For the remainder of the text, we will refer to our AI-enhanced SRD protocol simply as SRD.

Our experiment, schematically illustrated in Fig.~\ref{fig:1}(a), consists of strontium optical clock qubits ($\ket{g} = 5s^2 \, \singszero$, $\ket{e} = 5s5p \, \trippzero$) in an optical lattice of spacing $a_{\rm lat} \approx 575\, \si{\nano\meter}$~\cite{eckner2023realizing, cao2024multi}. 
$N_{\rm tot}$ atoms are arranged into rectangular ensembles of dimension $L_x \times L_y \geq N_{\rm tot}$, with a typical filling fraction of around 97\% and $2a_{\rm lat}$ or $3a_{\rm lat}$ spacing.
The clock state is coupled to the Rydberg state $\ket{r} = 5s47s \, \tripsone$.
To suppress non-adiabatic Rydberg excitations, dressing pulses are performed by ramping up to and down from a Rabi frequency $\Omega_r = 2 \pi \times 3\, \si{\mega \hertz}$ in $0.2 \, \si{\micro \second}$; simultaneously, the detuning $\Delta$ is ramped between $4 \Omega_r$ and $2 \Omega_r$, yielding a maximum dressing parameter $\beta = \Omega_r/2\Delta = 0.25$.
Conceptually, $\beta$ characterizes the admixture of Rydberg character in the dressed state $\ket{\tilde{e}} \sim \ket{e} + \beta \ket{r}$ which $\ket{e}$ is adiabatically connected to.
BBR primarily drives transitions from $\ket{r}$ to nearby contaminant states $\ket{r'}$ in the $5s np \, \trippzot$ Rydberg series.
A $\lambda \approx 407.89 \, \si{\nano\meter}$ AI laser~\cite{supp} is used to excite the core electron to a $5p_{3/2} nl$ AI state; notationally, we distinguish between the $\ket{r^*} = 5p_{3/2} 47s_{1/2}$ state excited from the target Rydberg state $\ket{r}$, and other $\ket{r'^{*}}$ states excited from all other BBR populated states $\ket{r'}$ [see Fig.~\ref{fig:1}(a) inset].

The core result of this work is shown in Fig.~\ref{fig:1}(c). 
The experiment starts by preparing a $10\times10$ atom array in the clock state $\ket{e}$.
We probe the survival over time in $\ket{e}$ after Rydberg dressing for two protocols.
For CRD, we apply a single ($M=1$) dressing pulse while varying the pulse duration $t_p$.
For SRD, we apply a variable number $M$ of dressing pulses, each with a fixed $t_p \approx 0.23\, \si{\micro\second}$~\cite{supp}; in between dressing pulses, AI pulses of duration $t_{\rm AI} = 0.3 \,\si{\micro \second}$ are applied, followed by a wait time of $t_{\rm wait} = 2 \, \si{\micro \second}$.
The total dressing time for both cases is defined as $t_d = M t_p$.
Dressing times are plotted with respect to the calculated two-body interaction energy $V_2 = \beta^3 \hbar \Omega_r\left(1 + 2 N \beta^2 \right)^{-3/2}/2$, with $N$ the effective number of atoms in a Rydberg blockade radius~\cite{hines2023spin, supp}.
The data show an order of magnitude improvement in the dressing lifetime for SRD over CRD.
The survival histograms at specific times reveal the collective loss and its suppression by AI.
For CRD, a low survival tail rapidly develops even at relatively short times, and ultimately the distribution becomes multi-modal with peaks around full survival and full loss~\cite{zeiher2016many}.
This structure can be understood intuitively from a quantum jump perspective, with the different peaks corresponding to whether or not a BBR-induced jump (and subsequent collective loss) occurred.
In contrast, the SRD case shows a unimodal distribution which steadily decays.
By quickly removing the BBR-induced contaminants with AI, facilitated excitation is never triggered and essentially single-particle loss is recovered.

\begin{figure}
    \includegraphics[width = \columnwidth]{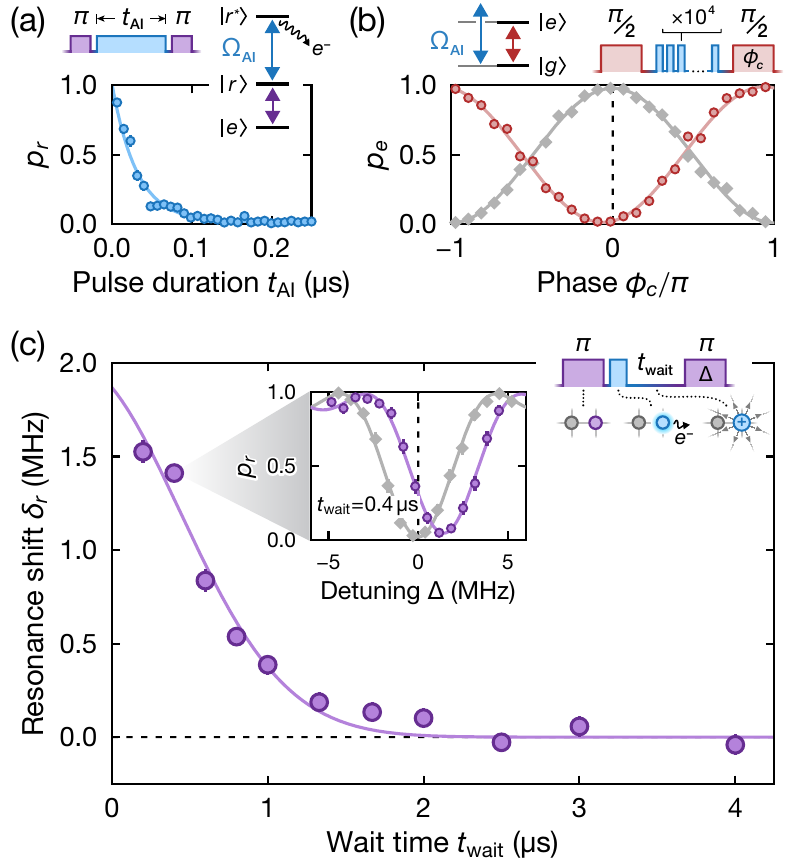}
    \caption{\label{fig:2}%
    \textbf{Characterizing autoionization of $\text{n=47}{{}}$ \textsuperscript{3}S\textsubscript{1} Rydberg atoms.}
    (a)~Duration of the AI process obtained by applying the AI laser for variable duration $t_{\rm AI}$.
    Here, $p_r$ (blue circles) corresponds to the detection probability in $\left| r\right\rangle$.
    (b)~Stark shift of the AI laser on the optical clock transition.
    Ramsey fringe (red circles) after application of $10^4$ $t_{\rm AI} = 0.2\,\si{\micro\second}$-long AI pulses during the dark time.
    Gray diamonds correspond to a reference measurement without the AI laser.
    The fitted fringe contrast is $C=\aicontrast$ [$\barecontrast$] with [without] applying the AI pulses and $p_e$  corresponds to the detection probability in $\left| e\right\rangle$.
    (c)~Transient Rydberg resonance shift (purple circles) for variable wait time $t_\mathrm{wait}$ after applying a $t_{\rm AI} =0.3 \, \si{\micro\second}$ AI pulse to a nearby Rydberg atom (initial separation $d=3a_{\rm lat}$), as illustrated in the top right schematic.
    Purple circles in the inset show an example spectroscopy measurement at $t_\mathrm{wait}= 0.4\,\si{\micro\second}$.
    Gray diamonds show a reference spectroscopy measurement.
    In panels~(a-c), pulse diagrams illustrate the experimental sequence with all atoms starting in $\ket{e}$ for (a), (c) and $\ket{g}$ for (b).
    Solid lines are numerical fits, and error bars indicate Clopper-Pearson confidence intervals for probabilities $p$ and fit errors for $\delta$.
    }
\end{figure}

To determine the potential improvement in SRD duty cycle, we characterize the AI rate in Fig.~\ref{fig:2}(a) by measuring the survival of atoms prepared in $\ket{r}$ after an AI pulse of duration $t_{\rm AI}$.
The data show a $1/e$ timescale of $\tau_{\rm AI} = \aitime$, which is three orders of magnitude faster than the relevant Rydberg state lifetimes~\cite{cao2024multi}.
The rate is currently limited by the AI laser intensity which determines the Rabi frequency $\Omega_{\rm AI}$ that the core transition $\ket{r} \leftrightarrow \ket{r^*}$ is driven at~\cite{supp}; the fundamental limit of roughly $\Gamma_{\rm AI}^{-1} \approx 15\,\si{\pico\second}$ is set by the intrinsic ionization rate $\Gamma_{\rm AI}$ of $\ket{r^*}$~\cite{xu1986sr,mukherjee2011many}.
The intrinsic rate varies with Rydberg principal number and angular momentum~\cite{cooke1978doubly,cooke1979calculation, yoshida2023autoionizing}, but will be similar for the states $\ket{r'^*}$ dominantly populated by BBR.
To ensure a high degree of contaminant removal, we use $t_{\rm AI}=0.2$--$0.3 \, \si{\micro \second}$ for SRD.

The AI leaves behind an ion~\footnote{The electron ejected by the AI departs rapidly due to its drastically smaller mass. The mass ratio between a $^{88}\mathrm{Sr}^+$ ion and an electron is of order $10^5$, and the departure time under constant acceleration scales inversely with the square root of the mass.} which substantially alters the nearby electric field environment, disturbing subsequent Rydberg excitation.
We investigate this effect in Fig.~\ref{fig:2}(c).
Isolated atom pairs are prepared into the Bell state $\left(\ket{er} + \ket{re}\right)/\sqrt{2}$ by performing a blockaded $\pi$-pulse~\cite{levine2018high, madjarov2020high}.
After AI of the $\ket{r}$ state, we perform Rydberg spectroscopy on the remaining $\ket{e}$ atom.
This atom samples the ion-generated field, shifting the resonance by $\delta_r$ (inset).
We observe $\delta_r$ decay for longer durations $t_{\rm wait}$ after the AI pulse as the ion moves further away.
The dominant mechanism for ejecting the ions is likely to be acceleration by stray electric fields, which we infer to be at the $360(20)\,\si{\milli\volt}/\si{\centi\meter}$ level~\cite{supp}.
For use in Rydberg dressing, we fix $t_{\rm wait} = 2\,\si{\micro\second}$, which is the dominant limitation for the currently achievable duty cycle.

\begin{figure*}
    \includegraphics[width = \textwidth]{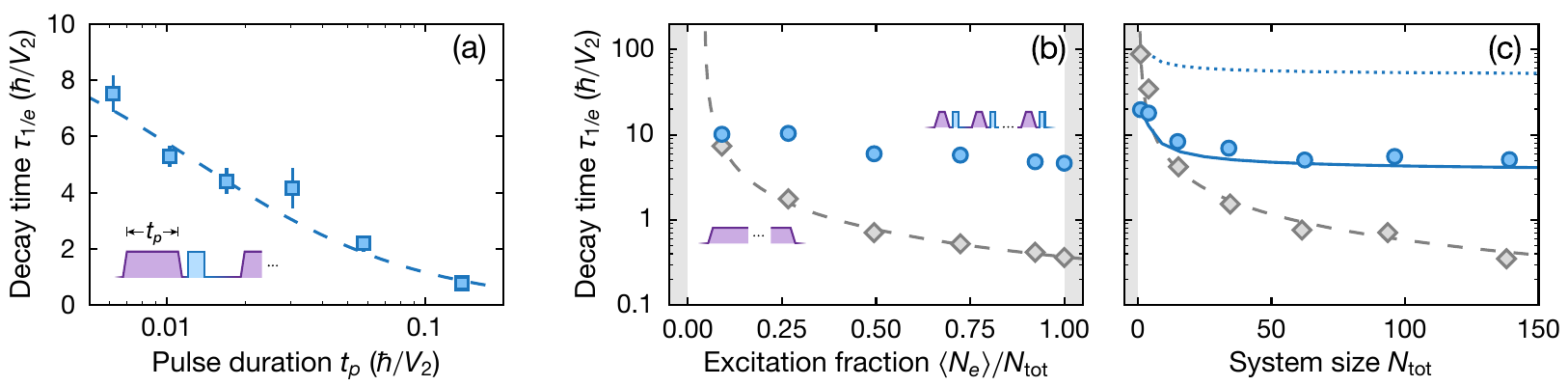}
    \caption{\label{fig:3}%
    \textbf{Benchmarking contaminant removal through autoionization laser pulses.}
    (a)~Decay time~$\tau_{1/e}$ (blue squares) for variable Rydberg dressing pulse duration $t_p$ in the SRD protocol, as illustrated in the bottom left inset pulse diagram.
    (b,c)~Decay time~$\tau_{1/e}$ for (b)~variable initial excitation fraction $\langle N_e \rangle / N_\mathrm{tot}$ in a constant $N_\mathrm{tot} = 10 \times 10$ array and (c)~variable system size $N_\mathrm{tot} = \sqrt{N_{\rm tot}} \times \sqrt{N_{\rm tot}}$ at constant $\langle N_e \rangle / N_\mathrm{tot}=1/2$.
    Blue circles (gray diamonds) correspond to SRD (CRD) protocols [see inset pulse diagrams in panel~(b)].
    Gray shaded regions indicate (experimentally) inaccessible regions.
    In panel (c), the solid blue line shows a constant $\tau_{1/e}(N_{\rm tot} = 1)$ [i.e. only scaled by the variation in $V_2(N_{\rm tot})$], and the dotted blue line shows the ideal Rydberg-decay-limited dressing lifetime.
    In all panels, dashed lines indicate numerical fits of $\tau_{1/e}^{-1}$ to a linear form.
    Each data point is extracted from a numerical fit and error bars indicate bootstrapped fit errors (partly smaller than marker size).
    }
\end{figure*}

In Fig.~\ref{fig:2}(b), we further show that exposure to the AI laser does not substantially degrade optical-clock-qubit coherence.
Employing a standard Ramsey interferometry protocol, we find that the fringe contrast is not reduced even after applying $10^4$ AI pulses during the dark time.
From a separate Ramsey measurement, we infer a differential clock-transition Stark shift of $\aishift$ due to the AI laser, consistent with expectation from the measured AI rate~\cite{supp}.
The phase shift per $1/e$ AI time, which is independent of AI laser intensity in the currently relevant regime $\Omega_{\rm AI} \ll \Gamma_{\rm AI}$, is \aishiftfom.

With the AI process well-characterized, we now return to assessing the Rydberg dressing lifetime improvement using SRD. 
While more sophisticated modeling for collective loss dynamics has been developed~\cite{aman2016trap,young2018dissipation}, we choose to simply characterize the dressing lifetime by a $1/e$ decay time $\tau_{1/e}$ extracted from an empirical fit of the survival $p_e$ over dressing time $t_d$~\cite{supp}.
We begin by investigating $\tau_{1/e}$ for SRD as a function of pulse duration $t_p$ (with fixed ramp time) in Fig.~\ref{fig:3}(a).
For these data, a $10 \times 10$ ensemble is initially prepared in the equal superposition $\left(\ket{g} + \ket{e}\right)/\sqrt{2}$ prior to dressing.
The decay time shows a monotonic decrease for longer $t_p$ as the per-pulse probability of BBR-induced decay and subsequent Rydberg facilitation is increased.
For the remaining results, we set $t_p$ to the lowest value shown, corresponding to an $\approx 8\%$ duty cycle for the largest systems explored.

To illustrate the suppression of the collective nature of the loss, we investigate $\tau_{1/e}$ as a function of density and system size in Figs.~\ref{fig:3}(b) and (c) respectively.
As a proxy for density, we vary the average initial excitation fraction $\langle N_e \rangle/N_{\rm tot} = \sin^2 \left( \theta /2 \right)$ by preparing each atom of a $10 \times 10$ array into the superposition $\cos \left( \theta/2 \right)\ket{g} + \sin \left( \theta/2 \right) \ket{e}$ with a clock $\theta$-pulse.
To vary the system size, we instead use arrays of variable size $N_{\rm tot} = \sqrt{N_{\rm tot}} \times \sqrt{N_{\rm tot}}$ with fixed $\theta = \pi/2$ ($\langle N_e \rangle/N_{\rm tot}=1/2$).
Indeed, we observe a markedly steeper decrease of the decay time with both excitation fraction and system size for CRD compared to SRD.
For the largest and densest systems, the lifetime improvement using SRD is an order of magnitude.

The SRD results are not yet at the fundamental limit $\tau_r/p_r$ set by the Rydberg state lifetime $\tau_r$ and dressing fraction $p_r$ [Fig.~\ref{fig:3}(c) dotted line].
One challenge is non-adiabaticity of the pulses. 
Indeed the faster single-atom ($N_{\rm tot} = 1$) decay for SRD compared to CRD appears to be explained by non-adiabaticity based on numerical simulations~\cite{supp}.
Interestingly, we find that the trend in $\tau_{1/e}$ with larger $N_{\rm tot}$ for SRD is roughly consistent with just the variation in $V_2$ at a fixed per-atom loss rate (solid line).
The degree to which non-adiabatic excitations affect $\tau_{\rm 1/e}$ for these larger systems is less clear, but non-adiabaticity certainly limits the minimum pulse duration which sets the residual collective loss within a single pulse [Fig.~\ref{fig:3}(a)].
Further improving the dressing lifetime remains an open challenge; nevertheless, the current results already demonstrate qualitatively improved scaling of dressing lifetimes for large systems by using SRD.

\begin{figure}
    \includegraphics[width=\columnwidth]{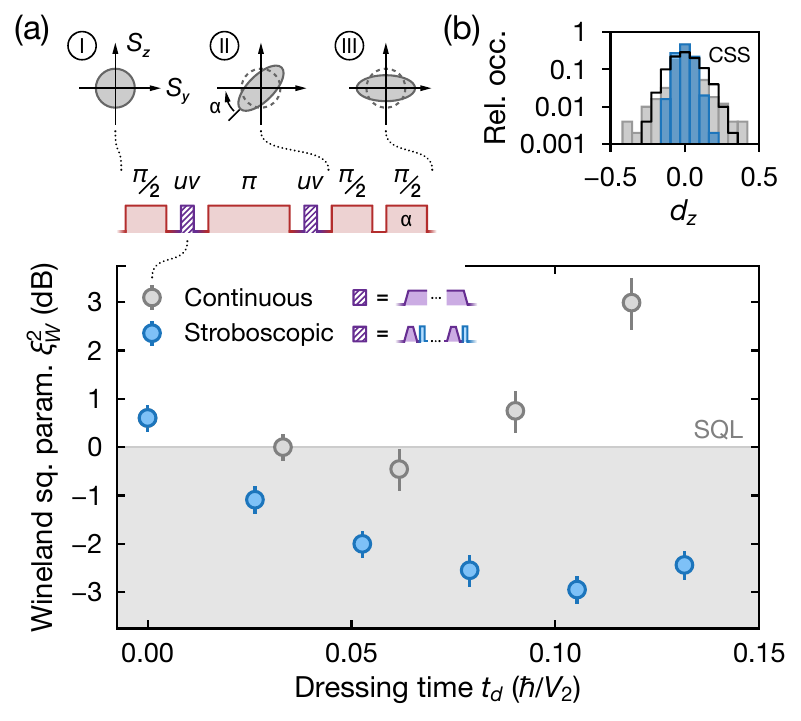}
    \caption{\label{fig:4}%
    \textbf{Preparing spin-squeezed states with autoionization-enhanced stroboscopic Rydberg dressing.}
    (a)~(top) Pulse sequence for the preparation of spin-squeezed states and schematic illustration of the quasi-probability on the Bloch sphere (2D projection shown) at different points (I)-(III) in the sequence.
    The pulses labelled ``uv'' correspond to either CRD or SRD (see legend).
    (bottom) Wineland squeezing parameter $\xi^2_W$ determined from variance and contrast measurements at variable dressing time $t_d$ for CRD (gray circles) and SRD (blue circles).
    Here, $\xi^2_W$ is determined from a differential measurement between two independent ensembles each with $N_\mathrm{tot}\approx 68$ atoms.
    Note that the $t_d = 0$ data corresponds to an unentangled CSS.
    SQL denotes the standard quantum limit.
    (b)~Histogram of the differential observable $d_z$ obtained for the variance measurements in panel~(a) at $t_d \approx 0.11\hbar/V_2$ for SRD (blue) and $t_d \approx 0.12\hbar/V_2$ for CRD (gray).
    Black lines correspond to a reference measurement with a CSS. 
    }
\end{figure}

To explicitly demonstrate the utility of our approach, we investigate spin-squeezing using SRD in Fig.~\ref{fig:4}.
Compared to an unentangled coherent spin state (CSS), spin-squeezed states feature reduced projection noise along a given spin quadrature to achieve quantum-enhanced measurement precision~\cite{wineland1992spin, wineland1994squeezed}.
Spin-squeezing is naturally generated by the soft-core Ising interactions of Rydberg dressing~\cite{gil2014spin, eckner2023realizing, hines2023spin}, which implement a finite-range version of the canonical one-axis twisting Hamiltonian~\cite{kitagawa1993squeezed}.
The spin-squeezing protocol is shown in the diagram of Fig.~\ref{fig:4}(a), following the procedure in Ref.~\cite{eckner2023realizing}.
We measure the differential observable $d_z = S_z^A/N_{\rm tot}^A - S_z^B/N_{\rm tot}^B$ between a pair $(A,B)$ of $5 \times 14$ ensembles to reject technical noise.

The measured distribution of $d_z$ after a dressing time $t_d \approx 0.11 \hbar/V_2$ is shown in Fig.~\ref{fig:4}(b). 
While collective loss leads to wide wings for CRD, the SRD shows a distribution which is strictly more narrow than the CSS ($t_d = 0$).
The metrological gain can be characterized by the Wineland squeezing parameter $\xi_W^2$~\cite{wineland1992spin,wineland1994squeezed}, which accounts for both the variance in $d_z$ as well as the spin vector length which we extract from a Ramsey fringe.
The improved gain is shown in the bottom panel of Fig.~\ref{fig:4}(a) where the SRD achieves around $3\,$dB below the standard quantum limit (SQL).
In contrast, the squeezing begins to degrade much earlier for CRD as the collective loss sets in, limiting the improvement to $<1\,$dB.
We note that the CRD results are significantly worse than realized in our previous work~\cite{eckner2023realizing}.
We are currently investigating the source of this discrepancy, but we emphasize that it is unlikely to be related to AI which is not used for CRD; given this and the near-SQL performance of the CSS, we suspect the disparity is related to the Rydberg laser.

In conclusion, we have demonstrated the use of AI for enhanced Rydberg dressing. 
By removing BBR-generated contaminants using AI in a SRD protocol, we achieved a 10-fold improvement in dressing lifetime for 144-atom systems while maintaining an order of magnitude larger duty cycle than previous work without AI~\cite{hines2023spin}. 
We showed that SRD is compatible with coherent operation of an optical clock qubit, and demonstrated improved spin squeezing at short dressing times.
The duty cycle could be further improved by incorporating fast-switching electrodes~\cite{loew2007apparatus} for rapid ion removal and increasing the AI laser intensity.
The extended lifetimes of our protocol open the door to longer-time, larger-scale quantum simulations of spin models, which can, for instance, enable improved spin-squeezing~\cite{kaubruegger2019variational,young2023enhancing} or exploration of exotic many-body physics~\cite{grusdt2013fractional,bijnen2015quantum,glaetzle2015designing}.
Alternatively, direct detection of the ionized contaminants could be a useful tool in erasure conversion schemes~\cite{wu2022erasure, scholl2023erasure, ma2023high}.

\begin{acknowledgments}
We acknowledge J.~Ye and his lab for the operation and provision of the silicon-crystalline-cavity-stabilized clock laser. 
We acknowledge stimulating discussions with S.~Hollerith, W.R.~Milner, R.~Potvliege, P.~Weckesser, J.T.~Young, J.~Zeiher.
The authors also wish to thank A.~Carroll and R.~Kaubruegger for careful readings of the manuscript and helpful comments.
This material is based upon work supported by the Army Research Office (W911NF-22-1-0104), the National Science Foundation QLCI (OMA-2016244), the National Science Foundation JILA-Physics Frontier Center (PHY-2317149), the U.S. Department of Energy, Office of Science, the National Quantum Information Science Research Centers, Quantum Systems Accelerator, and the National Institute of Standards and Technology. 
We also acknowledge funding from Lockheed Martin.
A.C. acknowledges support from the NSF Graduate Research Fellowship Program (Grant No. DGE2040434); W.J.E. acknowledges support from the NDSEG Fellowship; N.D.O. acknowledges support from the Alexander von Humboldt Foundation.
\end{acknowledgments}

\clearpage

\end{document}


\setstcolor{red}

\title{Supplementary Material for Autoionization-enhanced Rydberg dressing by fast contaminant removal}

\author{Alec Cao}
\author{Theodor \surname{Lukin Yelin}}
\author{William J. Eckner}
\author{Nelson \surname{Darkwah Oppong}}
\author{Adam M. Kaufman}
\affiliation{%
JILA, University of Colorado and National Institute of Standards and Technology,
and Department of Physics, University of Colorado, Boulder, Colorado 80309, USA
}%


\maketitle

\tableofcontents

\renewcommand{\thefigure}{S\arabic{figure}}
\renewcommand{\theequation}{S\arabic{equation}}

\section{Autoionization}

\subsection{Laser system and spectroscopy}

\label{sec:ailaser}

The autoionization (AI) transition is driven by an intracavity doubled VECSEL (Vexlum VALO SHG SF).
We typically work with 150--250$ \, \si{\milli\watt}$ out of the laser.
The wavelength is determined by a wavemeter (HighFinesse WS6-600) which is calibrated using light from a $689\,\si{\nano\meter}$ laser referenced to the intercombination transition $\singszero \leftrightarrow \trippone$ in \Sr~\cite{ido2005precision}.
The AI laser is focused through an acousto-optic modulator (AOM) (AA Opto Electronic MQ240-B40A0,2-UV) to perform fast pulses.
Around $30$--$40\,\si{\milli\watt}$ of light is delivered to the atoms via an optical fiber.

\begin{figure}
    \centering
    \includegraphics[width = \columnwidth]{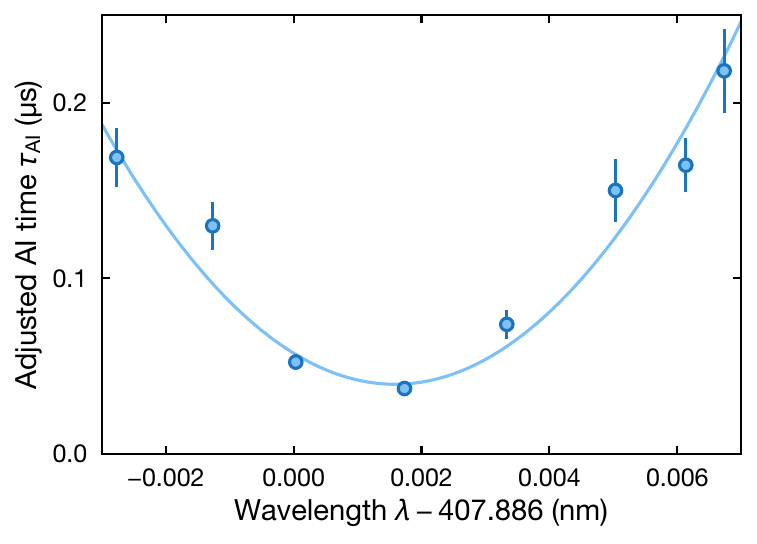}
    \caption{\textbf{Spectroscopy of the autoionization transition.}
    Measured exponential $1/e$ AI time $\tau_{\rm AI}$ as a function of the AI laser wavelength $\lambda$.
    $\tau_{\rm AI}$ is adjusted to compensate for variations in the laser power as the wavelength is changed.
    The line is a parabolic fit which extracts a transition wavelength of \aiwl.
    The $x$-axis of the plot is referenced to the transition wavelength of the $5s_{1/2} \leftrightarrow 5p_{3/2}$ transition in $\Sr^+$.}
    \label{fig:s1}
\end{figure}

In Fig.~\ref{fig:s1}, we refine previous spectroscopy results \cite{scholl2023erasure} to obtain a more precise determination of the AI transition wavelength. 
The spectroscopy is performed for the $5s47s \, \tripsone \leftrightarrow 5p_{3/2}47s_{1/2}$ transition.
We repeat the AI time $\tau_{\rm AI}$ measurement of Fig.~2(a) with different laser wavelengths.
The laser wavelength is tuned by adjusting the etalon temperature on the Vexlum source; to compensate for variations in the laser power as the wavelength was varied, the measured $\tau_{\rm AI}$ are adjusted proportionally.
A quadratic fit is used to find the center of the line at \aiwl, with the uncertainty indicating the fit error.
This measurement is taken in a 275$\,$G magnetic field with all optical trapping potentials turned off.
The wavelength of the relevant transition in $\Sr^+$ is $407.886\,\si{\nano\meter}$~\cite{moore1971atomic}.

\subsection{Clock transition light shift}

\label{sec:clockshift}

\begin{figure}
    \centering
    \includegraphics[width=\columnwidth]{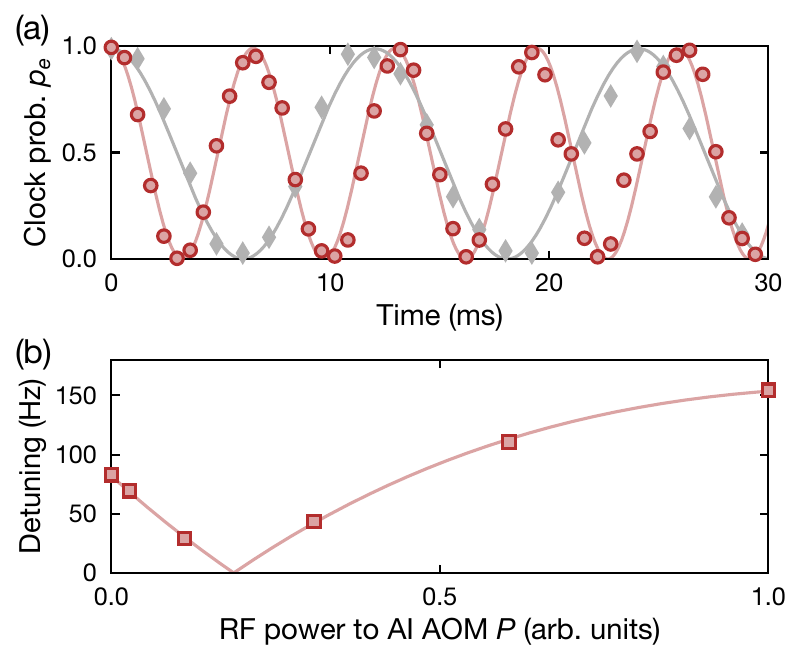}
    \caption{\textbf{AI laser-induced differential clock Stark shift.}
    (a) Clock state population $p_e$ as a function of Ramsey dark time.
    Red circles (gray diamonds) indicate data for which the AI laser was on (off) during the dark time.
    Solid lines indicate sinusoidal fits to extract the detunings during the dark time.
    (b) Fitted detuning magnitude for varying RF power $P$ sent to the AI laser AOM.
    The AI laser intensity $I$ varies sinusoidally with $\sqrt{P}$.
    The solid line is a linear fit of the detuning magnitude to $I$.}
    \label{fig:s2}
\end{figure}

Here we quantify the differential Stark shift induced on the optical clock transition $ \singszero \leftrightarrow \trippzero $ by the AI laser.
We perform a Ramsey interrogation on the clock transition with the atoms exposed to the AI laser continuously during the Ramsey dark time.
The clock state population as a function of Ramsey dark time is shown in Fig.~\ref{fig:s2}(a).
The fringe oscillates at a different frequency compared to the case without the AI laser where the detuning is set by the clock-laser probe shift.
The magnitude of the detuning is extracted using a sinusoidal fit.
In Fig.~\ref{fig:s2}(b), we repeat the measurement for varying AI laser intensity $I$.
Because we did not independently characterize $I$, we instead report these results with respect to the RF power $P$ sent to the AI laser AOM.
We fit the extracted detunings to the form $\abs{A + B \sin^2 \left( \frac{\pi}{2} \sqrt{\frac{P}{P_0}} \right)}$ (i.e. linear in $I$) based on the standard AOM first-order diffraction response, with $A$, $B$ and $P_0$ as fit parameters.
Based on this, we obtain a differential Stark shift magnitude of $h \abs{\delta_c} = h \times \aishift$ at the same intensity used for measuring the AI rate in Fig.~2(a).

\subsection{Theoretical autoionization rate}

\label{sec:airate}

Here we discuss the theoretical expectation for the AI rate. 
This rate is determined by two factors: the Rabi frequency $\Omega_{\rm AI}$ coupling the Rydberg and AI state, and the actual rate of ionization in the AI state $\Gamma_{\rm AI}$.
In principle, $\Gamma_{\rm AI}$ limits the maximum possible AI rate in the regime $\Omega_{\rm AI} \gg \Gamma_{\rm AI}$ where it would be possible to fully populate the AI state in a coherent manner.
In practice, given the natural linewidth of the transition $\Gamma_*$, the achievable laser intensity $I$ and the Rydberg principal number $n$ used, we operate far in the $\Omega_{\rm AI} \ll \Gamma_{\rm AI}$ regime.
In this regime, the AI rate of the Rydberg state can be approximated from Fermi's golden rule to be $ \approx \Omega_{\rm AI}^2/\Gamma_{\rm AI}$~\cite{cohen1998atom}.


Based off of spectroscopic measurements in the principal number range $n=10$--$21$, $\Gamma_{\rm AI}$ for the $5p_{3/2} ns_{1/2}$ states can be approximated as $\Gamma_{\rm AI} \approx 2\pi \times 9 \times 10^{14} \, \mathrm{Hz}/\nu^3$~\cite{xu1986sr, mukherjee2011many}, with $\nu = n - \delta$ the effective principal number and $\delta$ the quantum defect of the AI state.
For the $\tripsone$ Rydberg series, the quantum defect $\delta$ has been experimentally determined to high precision over the range $n=13$--$50$~\cite{couturier2019measurement}; the defect of the corresponding AI states has been measured to be within a couple percent of this for smaller $n$~\cite{xu1986sr}, though it has not been determined for the relevant $n=47$.
Taking $\delta \approx 3.37$~\cite{couturier2019measurement} gives $\Gamma_{\rm AI} \approx 2 \pi \times 10.8 \, \si{\giga\hertz}$ for $n=47$.
We note that for the $\trippzot$ series and corresponding AI states (the rates of which we have not directly measured but are relevant for contaminant removal), significantly less data is available for both $\Gamma_{\rm AI}$ and $\delta$.

Assuming an AI laser intensity $I$, the Rabi frequency is given by $\Omega_{\rm AI} = \Gamma_* \sqrt{I/2I_{\rm sat}}$ where $I_{\rm sat}$ is the saturation intensity.
The natural linewidth of the relevant $^{88}$Sr$^+$ transition is $\Gamma_* \approx 2 \pi \times 22.8 \, \si{\mega\hertz}$~\cite{gallagher1967oscillator}.
This determines the saturation intensity $I_{\rm sat} = \pi h c \Gamma_*/3 \lambda^3 \approx 43.9 \, \si{\milli\watt}/\si{\centi\meter}^2$.
The AI laser intensity can be inferred from the differential Stark shift; the two are related by $h \delta_c = \frac{\Delta \alpha}{2 c \epsilon_0} I$~\cite{grimm2000optical}.
At $\lambda = 407.8876 \, \si{\nano\meter}$, we calculate the differential scalar dynamic polarizability to be $\Delta \alpha/(2 c \epsilon_0 h) \approx 2.72(4) \times10^{-2} \, \si{\hertz} \,/\left( \si{\milli\watt}/\si{\centi\meter}^2 \right)$~\cite{safronova2024private}.
From the measured differential Stark shift in Section~\ref{sec:clockshift}, we infer $I/I_{\rm sat} \approx 2 \times 10^2$ and $\Omega_{\rm AI} \approx 2\pi \times 0.23 \, \si{\giga \hertz}$.
From this, we find an expected $1/e$ AI time of $\tau_{\rm AI} \approx \Gamma_{\rm AI}/\Omega_{\rm AI}^2 \approx 34 \, \si{\nano\second}$, which is in reasonable agreement with our Fig.~2(a) AI rate measurement.


\section{Ion-induced Rydberg Stark shift and stray fields}

\begin{figure}
    \centering
    \includegraphics[width=\columnwidth]{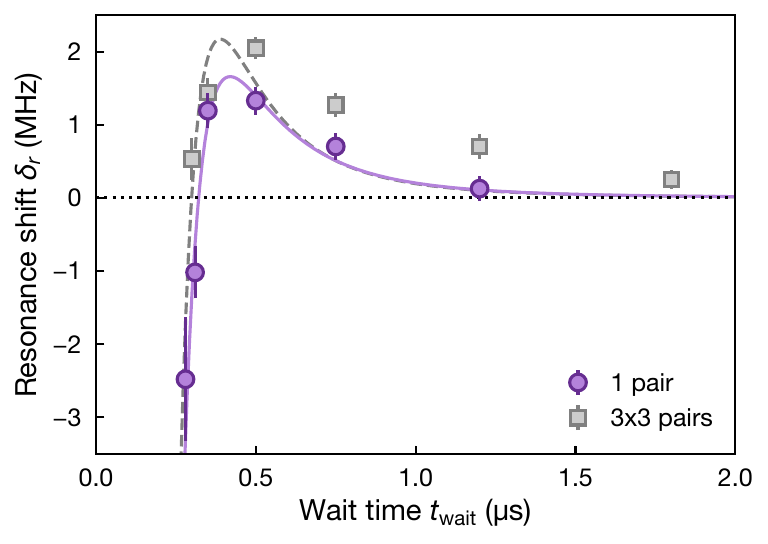}
    \caption{\textbf{Fitting Rydberg Stark shift data.}
    Rydberg resonance shift $\delta_r$ as a function of wait time $t_{\rm wait}$ after an AI pulse.
    The purple circles (gray squares) use a single atom pair ($3 \times 3$ array of atoms pairs).
    The solid purple (dashed gray) line are a corresponding fit to the Eq.~(\ref{eq:shift}) assuming constant acceleration under a uniform electric field.}
    \label{fig:s3}
\end{figure}

\label{sec:ionshift}

Here we discuss fitting the ion-induced Rydberg Stark shift, such as shown in Fig.~2(c), to a simple model for the motion of the ion in a stray field $\Estray$, which allows us to estimate the magnitude of the electric field in our system.
Let $\Eion$ denote the Coulomb field from the ion, and $\Etot = \Eion + \Estray$ denote the total field.
Let the atom being spectroscopically probed be located at the origin.
The measured shift $\delta_r$ is then
\begin{align}
    h \delta_r = - \frac{1}{2} \alpha_r \left( \abs{\Etot(0)}^2 - \abs{\Estray(0)}^2 \right).
\end{align}
Here, $\alpha_r \approx \alphar$ is the static polarizability of the $5s47s \tripsone$ Rydberg state, which we calculate using ARC~\cite{sibalic2017arc, robertson2021arc}.
For simplicity, we assume the trajectory of the ion is solely determined by a time-independent $\Estray$ oriented parallel to the $275\,$G bias magnetic field, which we denote the $z$-axis; this restricts the motion of the ion to that single direction.
The initial ion separation is $x_0 = 3 a_{\rm lat} \approx 1.7\,\si{\micro\meter}$, which we orient along the $x$-axis.
At a later separation $r = \sqrt{x_0^2 + z^2}$ between atom and ion, the Coulomb field at the atom is given by $\abs{\Eion(0)} = k e/r^2$.
The shift is then
\begin{align}
    h \delta_r = - \frac{1}{2} \alpha_r \left[ \frac{(k e)^2}{r^4} - \frac{2 \abs{\Estray(0)} k e z}{r^3}\right].
    \label{eq:shift}
\end{align}
Since $\alpha_r > 0$, the first term leads to large negative shifts at short times ($\abs{\Etot} > \abs{\Estray}$) while the second term dominates for longer times and yields positive shifts ($\abs{\Etot} < \abs{\Estray}$).
The sign of the observed shift $\delta_r > 0$ in Fig.~2(c) implies that we are primarily in the latter regime for all $t_{\rm wait}$ shown.
Note that the ion can already travel a significant amount during the AI pulse.
By assuming a form for $\Estray$, the trajectory $z(t)$ can be calculated and used to determine $\delta_r(t)$.
For a uniform $\abs{\Estray(z)} = E_0$, the ion trajectory is $z(t) = q E_0 t^2/2 m $.
The two terms in Eq.~(\ref{eq:shift}) then decay approximately as $(t/\tau_i)^{-8}$ and $(t/\tau_i)^{-4}$ respectively, with $\tau_i = \sqrt{2 m x_0/q E_0}$.

In Fig.~\ref{fig:s3}, we show data fitted to this constant acceleration, uniform field model.
We take the field strength $E_0$ and an offset time $t_0$ (such that $t = t_{\rm wait} + t_0$) as fit parameters.
These data were taken with a higher Rydberg Rabi frequency $\Omega_r \approx 2 \pi \times 5 \, \si{\mega\hertz}$ and faster AI time $\tau_{\rm AI} \approx 15\,\si{\nano\second}$ than the data presented in Fig.~2(c); faster pulses reduce the amount of averaging over the ion's motion, allowing us to measure the change in sign predicted in Fig.~(\ref{eq:shift}).
We find that constant acceleration model fits poorly for data taken using a $3\times3$ array of atom pairs, as was used in Fig.~2(c).
In particular, the data take much longer to decay, suggesting that the strength of the field reduces away from the initial location of the atoms.
We attribute this to ion-ion repulsion, which is non-negligible at the minimum inter-pair separation of $18a_{\rm lat} \approx 10\,\si{\micro\meter}$.
In contrast, the model fits quite well when using only a single atom pair, yielding a stray field strength of $E_0\approx 360(20) \, \si{\milli\volt}/\si{\centi\meter}$.

\section{Rydberg dressing}

\subsection{Saturation in the strong dressing regime}

\label{sec:saturation}

\begin{figure*}
    \centering
    \includegraphics[width = \textwidth]{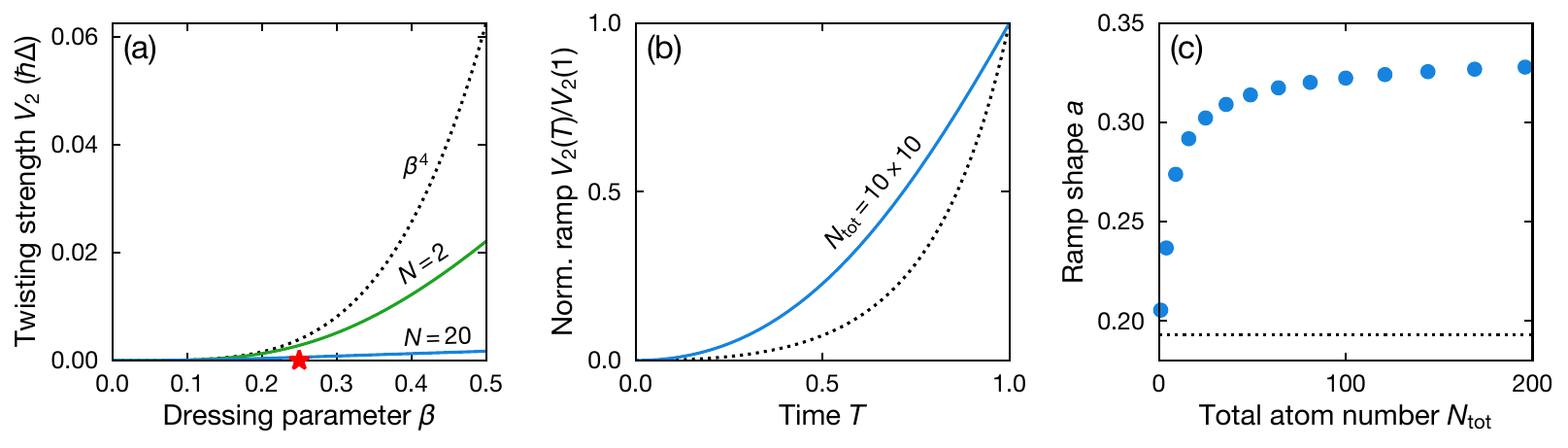}
    \caption{
    \textbf{Strong dressing saturation effects and effective pulse duration.}
    (a) Twisting strength $V_2$ [see Eq.~(\ref{eq:V2})] as a function of dressing parameter $\beta$ for $N=2$ (green) and $N=20$ (blue). 
    The black dotted line shows the $V_2/\hbar \Delta = \beta^4$ theory expected for weak dressing $\sqrt{N} \beta \ll 1$.
    The red star indicates the maximum $\beta = 0.25$ used for all results.
    (b) Normalized ramp of the twisting strength $V_2(T)/V_2(1)$ over time $T \in [0,1]$ during the ramp.
    Blue line shows the calculated ramp for the $N_{\rm tot} = 10 \times 10$ system used in the main text.
    The black dotted line shows the weak dressing limit with $V_2(T) = \beta^3(T) \hbar \Omega_r(T)/2$.
    (c) Ramp shape parameter $a$ for varying system size $N_{\rm tot} = \sqrt{N_{\rm tot}} \times \sqrt{N_{\rm tot}}$.
    The black dotted line shows the sytem size independent weak dressing limit.}
    \label{fig:s4}
\end{figure*}

Since we work at a relatively large dressing parameter of $\beta = 0.25$ and sufficiently dense ensembles, saturation effects can substantially alter the relevant energy timescales compared to the standard weak dressing expectation~\cite{henkel2010three, gil2014spin}. 
These effects have been discussed before in Refs.~\cite{balewski2014rydberg, hines2023spin}, and here we review them in order to quantitatively assess the measured dressing lifetimes.

We follow the formalism described in Ref.~\cite{hines2023spin}.
We first consider a fully blockaded $N$-atom system driven with Rabi frequency $\Omega_r$, detuning $\Delta$ and dressing parameter $\beta = \Omega_r/2\Delta$.
Due to the blockade, each Dicke state $\ket{N_e}$, with $0 \leq N_e \leq N$ the number of atoms in $\ket{e}$, evolves as an independent two-level system, coupling to a state with a single shared Rydberg excitation (except for $N_e=0$). The Hamiltonian describing the evolution of the $N_e$ sector is
\begin{align}
    \frac{H(N_e)}{\hbar \Delta} = \begin{pmatrix}
        0 & \sqrt{N_e} \beta \\ \sqrt{N_e} \beta & 1
    \end{pmatrix}.
    \label{eq:HNe}
\end{align}
The ground state energy of $H$ is given by
\begin{align}
    \frac{E(N_e)}{\hbar \Delta} = \frac{1}{2} \left( 1 - \sqrt{1 + 4 N_e \beta^2} \right).
\end{align}
We would like to obtain an effective Hamiltonian $H_{\rm eff}$ on a dressed version of the $\lbrace \ket{g}, \ket{e} \rbrace$ subspace.
Assuming we adiabatically follow the ground state of $H(N_e)$ for each $N_e$, $H_{\rm eff}$ will simply be a diagonal matrix of the values $E(N_e)$.
Alternatively, it can be expanded in a power series of the collective $N$-atom $S_z$ operator.
That is
\begin{align}
    H_{\rm eff} = \sum_{N_e=0}^N E(N_e) \ket{N_e} \bra{N_e} = \sum_{k=0}^{\infty} V_k S_z^k.
\end{align}
In general, we have $S_z = N_e - N/2$.
From this, we can extract the coefficients $V_k$ for $k>0$ as 
\begin{align}
    V_k = \frac{1}{k!} \pdv[k]{E(N_e)}{N_e} \bigg\rvert_{N_e = N/2} = - \frac{\left( \frac{1}{2} \right)_k }{2k!} \frac{ \hbar \Delta \left(4 \beta^2 \right)^k}{\left( 1 + 2 N \beta^2 \right)^{k-1/2}}.
\end{align}
Here, $(x)_k$ denotes the falling factorial $(x)_k = \prod_{l=0}^{k-1} (x-l)$.
Evaluating this explicitly for $V_2$ yields~\cite{hines2023spin} 
\begin{align}
    V_2 = \frac{\beta^3 \hbar \Omega_r}{2 \left(1 + 2 N \beta^2 \right)^{3/2}}.
    \label{eq:V2}
\end{align}
$V_2$ represents the twisting strength of the standard one-axis twisting Hamiltonian $H_{\rm OAT} = V_2 S_z^2$~\cite{kitagawa1993squeezed}; it is also closely related to the leading order contribution to the two-body interaction energy, and thus we choose to compare all dressing times in the main text with respect to the time-scale $\hbar/V_2$.
Note that $V_2$ decreases monotonically with $N$ for fixed laser parameters, and in particular faster than the Rydberg dressing fraction $p_r$ (see Section~\ref{sec:lifetimelimit}).
This is the underlying reason why the dressing decay time, in these units, decreases with $N_{\rm tot}$ in Fig.~3(c) for stroboscopic Rydberg dressing (SRD) despite the suppression of collective loss.

In practice, replacing $N$ with $N_{\rm tot}$ in Eq.~(\ref{eq:V2}) is a poor approximation as the system is only fully blockaded for the smallest few system sizes.
To account for this, we consider the blockade radius~\cite{balewski2014rydberg}
\begin{align}
    R_c = \left(\frac{C_6}{\hbar \sqrt{\Omega^2 + 2 \Delta^2}} \right)^{1/6}.
\end{align}
This form interpolates between the resonant case $(C_6/\hbar\Omega)^{1/6}$ and the weak dressing case $(C_6/2\hbar\Delta)^{1/6}$, and $R_c$
characterizes the range of the two-body soft-core potential $V(r) = V(0)/[1 + (r/R_c)^6]$ which describes the Ising interaction energy for two atoms separated by a distance $r$.
We note that $V(0)$ is also well-approximated by $2 V_2$ evaluated at $N=2$, particularly in the weak-dressing limit.
We compute an effective $V_2$ for larger systems by replacing $N$ in Eq.~(\ref{eq:V2}) with the number of atoms within a radius $R_c$.
To account for boundary effects, we compute this number explicitly for each site in a given array configuration and average over all sites in the array.
For the largest $12\times12$ system, we find $N \approx 17.5$.
The saturation of $V_2$ is shown in Fig.~\ref{fig:s4}(a).

\subsection{Effective pulse duration}

\label{sec:tp}

The SRD protocol requires short pulses to minimize the probability of collective loss within a single pulse, as shown in Fig.~3(a).
Combining this requirement with adiabaticity considerations can result in a large fraction of the dressing pulse duration needing to be spent during the ramps as opposed to holding at a fixed value.
Because of this, the effective amount of dressing after $M$-pulses of SRD can be significantly less than for a single pulse of $M$-times the duration in continuous Rydberg dressing (CRD).
To fairly compare the two, we define an effective pulse duration $t_p = t_{\rm hold} + 2 a t_{\rm ramp}$, with $t_{\rm ramp}$ and $t_{\rm hold}$ the duration of the ramp and hold respectively. 
Here, $0 \leq a \leq 1$ characterizes the ``ramp shape" as the relevant parameter is ramped to a maximum value during the hold; the factor of 2 accounts for ramping both up and down.
As discussed in Section~\ref{sec:saturation}, we take $V_2$ to be the relevant parameter, though we note that other choices could be considered in certain instances (for example, see Section~\ref{sec:adiabatic}).
The Rabi frequency and detuning ramps approximately take the form
\begin{align}
    \Omega_r(T) = \Omega_r \sqrt{T}, \,\,\, \Delta(T) = \Delta_i + (\Delta_f - \Delta_i) T.
\end{align}
Here $\Omega_r = 2\pi \times 3 \, \si{\mega\hertz}$, $\Delta_i = 4 \Omega_r$, $\Delta_f = 2 \Omega_r$  and $T = t/t_{\rm ramp} \in [0,1]$.
From these ramps and a given array configuration, we can compute $V_2(T)$.
Then $a$ is given simply by integrating $a = \int_0^1 \dd T \, V_2(T)/V_2(1)$.
The ramp for our $10 \times 10$ system is shown in Fig.~\ref{fig:s4}(b); in Fig.~\ref{fig:s4}(c), we show the computed values of $a$ as a function of $N_{\rm tot}$ used for the results in Fig.~3(c).

\subsection{Rydberg lifetime limit}

\label{sec:lifetimelimit}

Here we discuss the theoretical Rydberg-decay-limited dressing lifetime shown in Fig.~3(c).
In Ref.~\cite{cao2024multi}, $1/e$ exponential timescales to occupy states bright and dark to our detection protocol from the target $5s47s \tripsone$ Rydberg state were measured to be $\tau_r^{\rm dark} = 51(2) \, \si{\micro \second}$ and $\tau_r^{\rm bright} = 86(3) \si{\micro \second}$.
Since we employ the same detection protocol for these results, we choose to only consider $\tau_r^{\rm dark}$ here.
However, we note that photoionization of bright states by the Rydberg laser~\cite{scholl2023erasure} could contribute to our measured signal.
Compared to the Rydberg state lifetime, the single-particle dressing lifetime should be increased by a factor of the Rydberg state dressing fraction $p_r$.
For our parameters, $p_r$ is also strongly modified relative to the weak dressing prediction $\beta^2$.
From the Hamiltonian in Eq.~(\ref{eq:HNe}), we can compute the per-atom Rydberg fraction to be
\begin{align}
    p_r(N_e) = \frac{1}{2N_e} \left( 1 - \frac{1}{\sqrt{1 + 4 N_e \beta^2}} \right).
    \label{eq:pr}
\end{align}
Note that unlike $V_2$, $p_r$ depends explicitly on the Dicke state number $N_e$.
Since we generally do not prepare Dicke states (except for $N_e = 0, N_{\rm tot}$), we simply make the approximate replacement $N_e \to \langle N_e \rangle = N \sin^2 \theta/2$ for atoms that have been prepared with a clock $\theta$-pulse.
$N$ is again replaced by the average number of atoms in a radius $R_c$ (see Section~\ref{sec:saturation}).
With these quantities, the ideal dressing loss rate is then $\gamma = p_r/\tau_r^{\rm dark}$.
The corresponding ideal lifetime $\gamma^{-1}$ is plotted in Fig.~3(c).

\subsection{Decay time fitting}

\label{sec:decaytime}

The following procedure is used to extract the dressing decay times $\tau_{1/e}$ shown throughout Fig.~3.
We perform numerical fits of the excited state survival $p_e$ over time to the empirical model
\begin{align}
    p_e(t_d) = p_0 \left( f e^{- \gamma t_d} + (1-f) e^{- \kappa t_d} \right).
    \label{eq:fitfunc}
\end{align}
Here, $p_0$, $f$ and $\kappa$ are fit parameters, and $\gamma$ is the theoretical Rydberg-decay-limited loss rate (see Section~\ref{sec:lifetimelimit}).
We stress that this model is purely empirical, and generally the collective loss dynamics should not be expected to be well-modeled by an exponential behavior.
Nevertheless, we found Eq.~(\ref{eq:fitfunc}) to capture the data well over the large range of conditions considered in Fig.~3; representative examples of the fit are displayed as solid lines in Fig.~1(c).
After obtaining the fit parameters, we then compute the decay time by solving Eq.~(\ref{eq:fitfunc}) numerically under the condition $p_e(\tau_{1/e}) = p_0/e$.
We perform the fits using the raw survival fraction from each shot of the experiment (as opposed to the average over multiple repetitions which is plotted), and error bars are obtained by bootstrap resampling using $10^3$ resamples.

\begin{figure}
    \centering
    \includegraphics[width=\columnwidth]{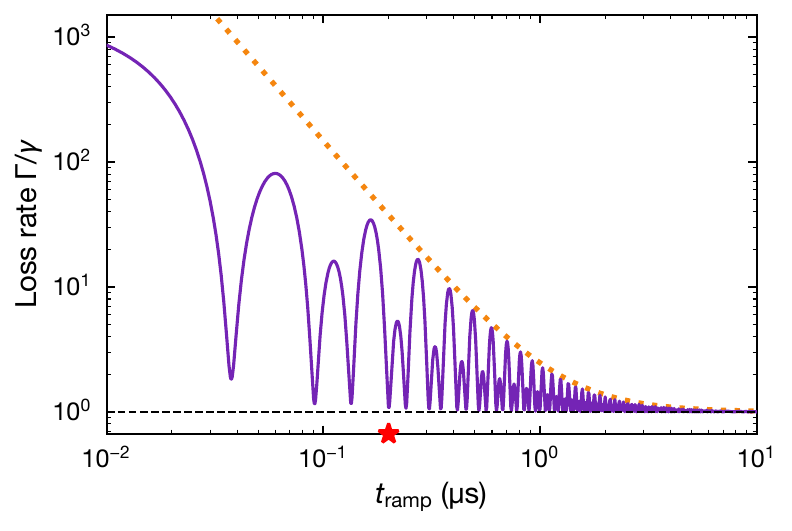}
    \caption{\textbf{Dressing pulse ramp adiabaticity.}
    Ratio of computed loss rate $\Gamma$ [see Eq.~(\ref{eq:Gamma})] to the ideal Rydberg-decay-limited rate $\gamma$ for varying ramp time $t_{\rm ramp}$ (purple solid line).
    The result is calculated from a single-particle simulation.
    The orange dotted line shows an empirical $\Gamma/\gamma-1 \propto t_{\rm ramp}^{-2}$ scaling which serves as a guide to the eye for the oscillation envelope.
    The horizontal black dashed line indicates $\Gamma/\gamma=1$, which is approached in the adiabatic limit.
    The red star indicates the $t_{\rm ramp} = 0.2\,\si{\micro\second}$ ramp time used in all results.
    }
    \label{fig:s5}
\end{figure}

\subsection{Pulse adiabaticity}

\label{sec:adiabatic}

As shown in Fig.~3(c), we measure a shorter single-atom dressing lifetime $\tau_{1/e}(N_{\rm tot} = 1 )$ for SRD compared to CRD; a similarly short lifetime was measured for the same SRD protocol without applying the AI laser (see Section~\ref{sec:srdnoai}).
One potential cause of this is non-adiabaticity of the dressing pulse ramps, the relative effect of which increases for more rapid strobing.
To investigate this hypothesis, we perform two-level numerical simulations of the pulse.
Explicitly, we use the Hamiltonian in Eq.~(\ref{eq:HNe}) for $N_e = 1$, but replace the excited state energy (lower right diagonal) by $1 \to 1 - i /(2 \Delta \tau_r^{\rm dark})$ to capture the Rydberg state lifetime.
From this, we compute the single-pulse evolution operator $U$ and obtain $U_{e,e} = \bra{e} U \ket{e}$ as the transition amplitude to return to the initial state.
Assuming the Rydberg state population is fully removed by AI pulses, the clock-state survival will then decay exponentially at a rate $\Gamma$ given by
\begin{align}
    \Gamma = - \frac{2 \ln \abs{U_{e,e}}}{t_p}.
    \label{eq:Gamma}
\end{align}
In an adiabatic regime, $\Gamma/\gamma$ should be constant as a function of pulse duration $t_p$.

The results for $\Gamma/\gamma$ as a function of the ramp time $t_{\rm ramp}$ are shown in Fig.~\ref{fig:s5}; the hold time is fixed to $t_{\rm hold} = 0.1 \, \si{\micro\second}$.
Restricted to reporting these simulation results, we calculate $a$ for the effective pulse duration $t_p$ based on the Rydberg probability $p_r$ (instead of $V_2$) such that $\Gamma/\gamma \to 1$ in the adiabatic limit, as seen for the largest $t_{\rm ramp}$.
The ramp time of $t_{\rm ramp} = 0.2\, \si{\micro \second}$ used in the main text is in a highly non-adiabatic regime with large oscillations in $\Gamma/\gamma$.
The simulations actually show a minima almost precisely at $t_{\rm ramp} = 0.2 \, \si{\micro \second}$, but we suspect that modeling of the experiment at the $0.01\,\si{\micro \second}$ level could be inaccurate, for instance due to the finite bandwidth of the AOM used to generate the pulses.
Note that while the CRD protocol uses the same ramp parameters, the relative contribution of non-adiabaticity is reduced by the longer $t_{\rm hold}$.
Further exploration is needed for the contribution of non-adiabaticity to the observed lifetimes of many-particle systems.

\subsection{Stroboscopic dressing without autoionization}

\label{sec:srdnoai}

\begin{figure}
    \centering
    \includegraphics[width=\columnwidth]{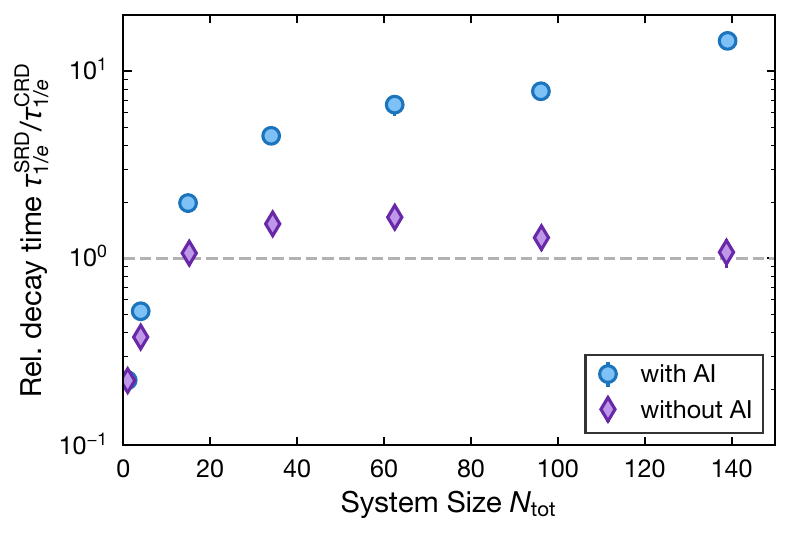}
    \caption{\textbf{Comparison of SRD with and without AI at a fixed duty cycle.}
    Dressing decay time $\tau_{1/e}$ of SRD relative to CRD for varying system size $N_{\rm tot}$.
    The blue circles (purple diamonds) indicate SRD with (without) applying the AI laser.
    The horizontal dashed line indicates a ratio of 1.}
    \label{fig:s6}
\end{figure}

To isolate the improvement in dressing decay time due solely to the AI at a given duty cycle, we additionally perform experiments using identical SRD conditions (i.e. dressing parameters and time between Rydberg pulses) but without applying the AI laser.
In Fig.~\ref{fig:s6}, we show the ratio of the SRD decay times with and without AI relative to CRD for varying atom number; note that data used for CRD and SRD with AI are identical to those shown in Fig.~3(c).
For small atom numbers without much collective loss, we find that the results with and without AI approach each other; as discussed in Section~\ref{sec:adiabatic}, both are worse than the CRD decay time due to non-adiabaticity of the pulse ramps.
For large systems, SRD without AI provides much less improvement than with AI at the given duty cycle.
In the future, it would be interesting to experimentally confirm whether or not the SRD without AI reaches the same performance as with AI for longer wait times between dressing pulses.